\documentclass[a4paper]{article}

\usepackage[includeheadfoot, left=3cm, right=3cm, top=2cm, bottom = 2cm]{geometry}
\usepackage{fancyhdr}
\usepackage{authblk}

\usepackage[utf8]{inputenc}
\usepackage[T1]{fontenc}
\usepackage{libertine}
\usepackage[UKenglish]{isodate}
\usepackage{tabularx}

\usepackage{graphicx} 
\usepackage{iftex}
\usepackage[xindy]{glossaries}

\newcommand{\eventb}{Event\nobreakdash-B\xspace}

\newacronym{ltl}{LTL}{Linear-time Temporal Logic}

\newacronym{ctl}{CTL}{Computation Tree Logic}

\newacronym{ptl}{PTL}{Probabilistic Temporal Logic}

\newacronym{pctl}{PCTL}{Probabilistic Computation Tree Logic}

\newacronym{pctl*}{PCTL*}{Probabilistic Computation Tree Logic*}

\newacronym{tctl}{TCTL}{Timed Computation Tree Logic}

\newacronym{stl}{STL}{Signal Temporal Logic}

\newacronym{ltl-x}{LTL-X}{a version of Linear Time Logic without the `next' operator}

\newacronym{mtl}{MTL}{Metric Temporal Logic}

\newacronym{iactlk-x}{IACTLK-X}{a fragment of the temporal-epistemic logic CTLK, without the `next' operator}

\newacronym{mucalc}{$\mu$-calculus}{a strict superset of other temporal logics}

\newacronym{fotl}{FOTL}{First-Order Temporal Logic}

\newacronym{bdi}{BDI}{Belief-Desire-Intention}

\newacronym{prs}{PRS}{Procedural Reasoning System}

\newacronym{are}{ARE}{Autonomous Reasoning Engine}

\newacronym{dmars}{dMARS}{distributed Multi-Agent Reasoning System}

\newacronym{fsm}{FSM}{Finite-State Machine}

\newacronym{pfsm}{PFSM}{Probabilistic Finite-State Machine}

\newacronym{apfsm}{APFSM}{\textit{Autonomous} Probabilistic Finite-State Machine}

\newacronym{pha}{PHA}{Probabilistic Hybrid Automata}

\newacronym{fsa}{FSA}{Finite-State Automata}

\newacronym{ta}{TA}{Timed Automata}

\newacronym{gspn}{GSPN}{Generalised Stochastic Petri Net}

\newacronym{mopn}{MOPN}{Marked Ordinary Petri Net}

\newacronym{ba}{BA}{B\"{u}chi Automata}

\newacronym{dtmc}{DTMC}{Discrete-Time Markov Chain}

\newacronym{pars}{PARS}{Process Algebra for Robot Schemas}

\newacronym{fsp}{FSP}{Finite State Processes}

\newacronym{piadl}{$\pi$ADL}{$\pi$ calculus combined with ADL}

\newacronym{csp}{CSP}{Communicating Sequential Processes}

\newacronym{ccs}{CCS}{Calculus of Communicating Systems}

\newacronym{wsccs}{WSCCS}{Weighted Synchronous CCS}

\newacronym{csp|b}{CSP$\|$B}{an integrated formalism combining CSP and B}

\newacronym{fdr}{FDR}{the Failures-Divergences Refinement checker}

\newacronym{jpf}{JPF}{Java PathFinder}

\newacronym{ajpf}{AJPF}{Agent Java PathFinder}

\newacronym{mcmas}{MCMAS}{Model Checker for Multi-Agent Systems}

\newacronym{ltlmop}{LTLMoP}{Linear Temporal Logic MissiOn Planning}

\newacronym{adl}{ADL}{Architecture Description Language}

\newacronym{aadl}{AADL}{Architecture Analysis and Design Language}

\newacronym{lts}{LTS}{Labelled-Transition System}

\newacronym{uml}{UML}{Unified Modelling Language}

\newacronym{sysml}{SySML}{Systems Modelling Language}

\newacronym{robotml}{RobotML}{Robot Modelling Language}

\newacronym{dsl}{DSL}{Domain Specific Language}

\newacronym{sct}{SCT}{Supervisory Control Theory}

\newacronym{ide}{IDE}{Integrated Development Environment}

\newacronym{mde}{MDE}{Model-Driven Engineering}

\newacronym{ai}{AI}{Artificial Intelligence}

\newacronym{fdir}{FDIR}{Fault Detection, Isolation and Recovery}

\newacronym{ifm}{iFM}{Integrated Formal Methods}

\newacronym{ros}{ROS}{Robot Operating System}

\newacronym{amcl}{AMCL}{Adaptive Monte Carlo Localisation}

\newacronym{dl}{\text{\upshape\textsf{d{\kern-0.05em}L}}}{differential dynamic logic}

\newacronym{vm}{VM}{Virtual Machine}

\newacronym{bip}{BIP}{Behaviour Interaction Priority}

\newacronym{fol}{FOL}{First-Order Logic}

\newacronym{owl}{OWL}{Web Ontology Language}

\newacronym{ria}{RIA}{Restore Invariant Approach}

\newacronym{ocl}{OCL}{Object Constraint Language}

\newacronym{sil}{SIL}{Safety Integrity Level}

\newacronym{rcl}{RCL}{ROS Contract Language}

\newacronym{rv}{RV}{Runtime Verification}

\newacronym{rml}{RML}{Runtime Monitoring Language}

\newacronym{sdlc}{SDLC}{Systems Development Life Cycle}
\usepackage{fretish}
\usepackage{amsfonts}
\usepackage{mathtools}
\usepackage{longtable}
\usepackage{array}
\usepackage{amssymb}
\usepackage{appendix}
\usepackage[hidelinks]{hyperref}
\usepackage{dafny}
\usepackage{bsymb,newestb2latex}
\usepackage{tikz}
\usetikzlibrary{arrows,automata,positioning}

\usepackage{paralist}
\usepackage{url}

\usepackage{orcidlink}

\usepackage[
    type={CC},
    modifier={by},
    version={4.0},
]{doclicense}

\fancypagestyle{firststyle}
{  
   \fancyhf{}
   \fancyfoot[L]{
   \begin{minipage}{\textwidth}
   \centering {\footnotesize \doclicenseThis }
   \thepage 
\end{minipage}  
}
 
}

\title{Adventures in FRET and Specification\thanks{This work was partially supported by the Royal Academy of Engineering and EPSRC grant EP/Y001532/1, as well as Maynooth University's Hume Doctoral Award.}}

\author[1]{Marie Farrell}
\author[2]{Matt Luckcuck}
\author[3]{Rosemary Monahan}
\author[1]{Conor Reynolds}
\author[3]{Ois\'{i}n Sheridan}

\affil[1]{Department of Computer Science, The University of Manchester, Manchester, UK}
\affil[2]{School of Computer Science, University of Nottingham, Nottingham, UK}
\affil[3]{Department of Computer Science, Maynooth University/Hamilton Institute, Maynooth, Ireland}

\begin{document}

\maketitle
\thispagestyle{firststyle}
\begin{abstract}
    This paper gives an overview of previous work in which the authors used NASA's Formal Requirement Elicitation Tool (FRET) to formalise requirements. We discuss four case studies where we used FRET to capture the system's requirements. These formalised requirements subsequently guided the case study specifications in a combination of formal paradigms. For each case study we summarise insights gained during this process, exploring the expressiveness and the potential interoperability of these approaches.  Our experience confirms FRET's suitability as a framework for the elicitation and understanding of requirements and for providing traceability from requirements to specification.
\end{abstract}

\section{Introduction}
\label{sec:introduction}
Knowing which properties to specify and subsequently verify is one of the biggest bottlenecks in the use of Formal Methods \cite{rozier2016specification}. Requirements are often written in natural-language, which can be ambiguous and is not generally amenable to formal verification. As a result, requirements elicitation and formalisation becomes even more important for understanding what to specify and verify. 

 Tools such as NASA's Formal Requirements Elicitation Tool (FRET) have been developed to bridge the gap between natural-language requirements and the logics normally used for formal specification \cite{giannakopoulou2020formal}. FRET provides a structured natural-language, called \fretish, from which temporal logic specifications and other verification conditions can be derived. 

We have used FRET in many use cases to support both elicitation and formalisation of requirements. In this paper, we provide an overview of four of these case studies and discuss how the \fretish requirements were elicited and subsequently used for specifying properties to be verified in various formal methods. These case studies are an aircraft engine software controller \cite{farrellfretting2022}, a mechanical lung ventilator \cite{abz2024}, a rover carrying out an inspection task \cite{bourbouhintegrating} and an algorithm for autonomously grasping spent rocket stages \cite{farrellformal2022}.

We outline the relevant background material in \S\ref{sec:background}. Next, we describe the four case studies that we explore in this paper (\S\ref{sec:casestudies}). \S\ref{subsec:analysis} provides metrics about our \fretish requirement sets, showing the structure of requirements and how to express them in  \fretish. We discuss how we used these requirements in other formal methods (model-checking, theorem proving and runtime verification approaches), encouraging a framework for interoperability. We provide a brief discussion in \S\ref{sec:discussion}. Finally, \S\ref{sec:conclude} concludes.

\section{Background}
\label{sec:background}
We provide an overview of the various formal methods used in the case studies reviewed in this paper: FRET, Dafny, Event-B and ROSMonitoring.

\subsection{FRET}
\label{sec:fretbg}

The \gls{fret} is an open-source tool that enables developers to write and formalise system requirements~\cite{giannakopoulouformal2020}. \gls{fret} requirements are written in a structured natural-language called \fretish, built from five fields: \fretishComponentsSmall{}.

\begin{figure}[t]
 \centering
 \includegraphics[width=0.9\textwidth]{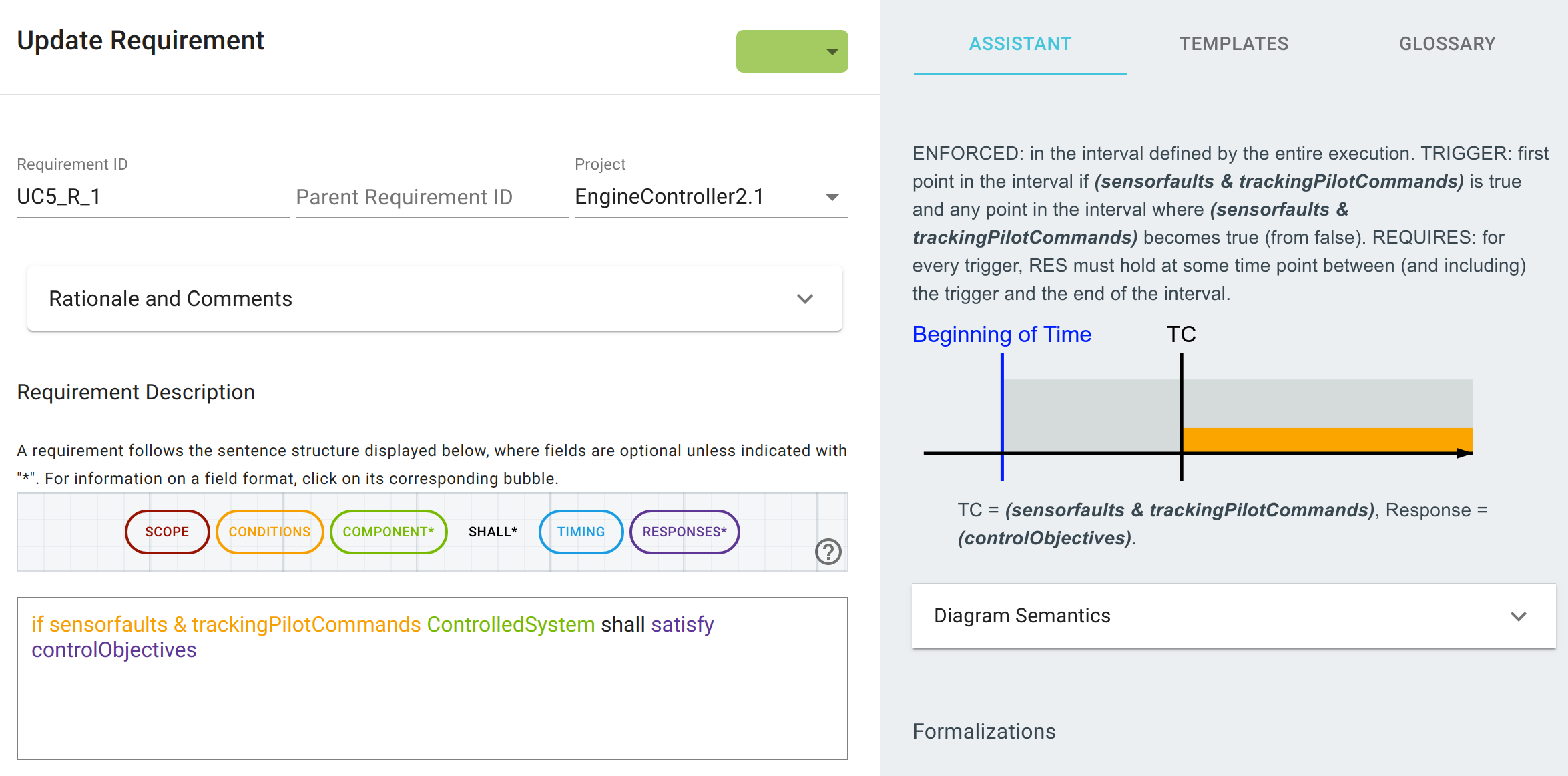}
 \caption{Screenshot of \gls{fret} showing the \fretish{} version of requirement UC5\_R\_1 from the Aircraft Engine Controller case study (\S\ref{subsec:enginecontroller}). Each requirement has an ID (top left), and a diagrammatic semantics (right) is generated by the tool to help the user to understand the meaning behind the \fretish{} requirement \cite{farrellfretting2022}.}
 \label{fig:fretR1}
\end{figure}

The \Component{} and \ResponseF{} fields are mandatory (along with the ``shall'' keyword) for all \fretish requirements; \Scope{},  \ConditionF{} and \Timing{} are optional.
Using \fretish, users can express requirements for individual \Component{}s that pertain to a particular \Scope{} under some \ConditionF{} where the \ResponseF{} (expected behaviour) can be specific to a defined \Timing{}. Fig.~\ref{fig:fretR1} is a screenshot of \gls{fret}'s `Update Requirement' dialogue, with the \fretish{} version of a requirement 
from the Aircraft Engine Controller case study, discussed in Section~\ref{subsec:enginecontroller}. 

\gls{fret} provides automated translations from \fretish requirements to CoCoSpec contracts \cite{champion2016cocospec}, which can be verified with the Kind2 model checker, and Copilot runtime monitors \cite{perez2022automated}. There is no automated support for translating between \fretish requirements and theorem proving approaches like \eventb, though prior work shows proof-based methods are useful for verifying requirements that are difficult to verify using the supported CoCoSpec approach \cite{bourbouhintegrating}. Other requirements tools include D-Risq's Kapture tool\footnote{Kapture: \url{https://www.drisq.com/product-kapture}}, IBM's DOORS\footnote{DOORS: \url{www.ibm.com/products/requirements-management?mhsrc=ibmsearch_a&mhq=DOORS}}, and the open-source tool Doorstop\footnote{Doorstop: \url{https://github.com/doorstop-dev/doorstop}}. \gls{fret} was chosen because it is a free, open-source platform that is under active development.

\subsection{Event-B}
\label{sec:ebbg}

\eventb is a state-based formal method that has been used in the development of safety-critical systems in a variety of sectors, including rail \cite{kiss2016developing}, aerospace \cite{mammar2017modeling} and medical \cite{hoang2016validating}. The \eventb language is based on set theory and first-order logic \cite{abrial_modeling_2010}. Tool support is provided by the Rodin Platform \cite{abrial2010rodin}. \eventb, like FRET, is open-source. \eventb specifications comprise \textit{machines} and \textit{contexts}. A machine specifies dynamic behaviour via variants, invariants, and events. Static behaviour is  specified in contexts,  using carrier sets, constants, and axioms.

\eventb supports formal refinement, enabling users to gradually add more detail to their model and discharge proofs of correctness at each stage \cite{schneider_behavioural_2014}. Theorem proving is supported in Rodin with most of the generated proof obligations discharged automatically, although some may need manual interaction with Rodin's provers. The semantics of \eventb models is often thought of in terms of the generated proof obligations \cite{hallerstede_purpose_2008}, and a detailed formal semantics for the \eventb language itself is given in \cite{farrell2022building}.

\begin{figure}[t]
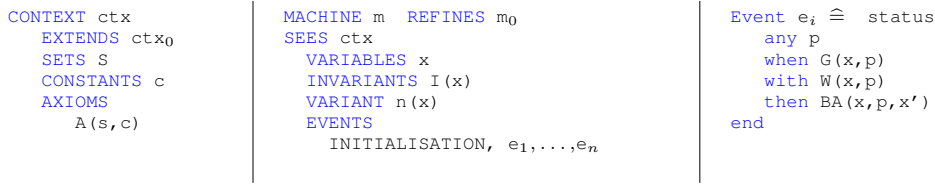

\begin{minipage}[t]{0.23\textwidth}
\begin{programsc}[3ex]*
\CONTEXT{ctx} 
   \EXTENDS{ctx$_0$}
   \SETS S
   \CONSTANTS c
   \AXIOMS  
      A(s,c)    
\end{programsc}
\end{minipage}
\rule[-16ex]{.1pt}{16ex}
\begin{minipage}[t]{0.38\textwidth}
\begin{programsc}[3ex]*
\MACHINE{m}  \REFINES{m$_0$}
\SEES{ctx}
  \VARIABLES x
  \INVARIANTS I(x) 
  \VARIANT n(x)
  \EVENTS  
    \texttt{INITIALISATION}, e$_1,\ldots,$e$_n$  
\end{programsc}
\end{minipage}
\rule[-16ex]{.1pt}{16ex}
\begin{minipage}[t]{0.34\textwidth}
\begin{programsc}[3ex]*
\EVT{e$_i$}\texttt{ status}
   \bkw{any} p
   \bkw{when} G(x,p)
   \bkw{with} W(x,p)
   \bkw{then} BA(x,p,x')
\bkw{end}\end{programsc}  

\end{minipage}

    \caption{Event-B specifications comprise contexts (static aspects) and machines (dynamic behaviour). Machines contain events that specify state changes.}
    \label{fig:ebsyntax}
\end{figure}

\subsection{Dafny}
\label{sec:dafnybg}

Dafny is a formal verification system that is used in the static verification of functional program correctness. Users provide specification constructs e.g. pre-/post-conditions, loop invariants and variants \cite{leino2010dafny}. Programs are translated into the Boogie intermediate verification language \cite{barnett2005boogie} and then the Z3 automated theorem prover discharges the associated proof obligations \cite{de2008z3}. 

Fig. \ref{fig:basicdafny} shows the basic structure of a Dafny method. The {\small\textbf{\texttt{requires}}} keyword (line 2) indicates the method's pre-condition, the {\small\textbf{\texttt{modifies}}} keyword (line 3) specifies which input variables the method is allowed to modify and the {\small\textbf{\texttt{ensures}}} keyword (line 4) captures the method's post-condition. The user specifies a loop {\small\textbf{\texttt{invariant}}} (line 7) which is used by the underlying SMT solver to reason about the loop's correctness and to prove that the post-condition is preserved, and the {\small\textbf{\texttt{decreases}}} clause (line 8) is a loop variant for proving loop termination.

\begin{figure}[t]
\begin{dafny}[numbers=left, stepnumber=1, frame=single]
method myMethod(x: int) returns (y: int)
requires ...
modifies ...
ensures ...
{
 while (i<x) //this is a comment
 invariant ...
 decreases ...
 {...}
}
\end{dafny} 
 \caption{The basic structure of a method with specification constructs in Dafny.}
 \label{fig:basicdafny}
\end{figure}

\subsection{ROSMonitoring}
\label{subsec:rosmonbg}

ROSMonitoring is a runtime verification tool for systems developed using the Robot Operating System (ROS)~\cite{Ferrando20a}. As shown in Fig.~\ref{fig:rosmon-pipeline}, a configuration file guides the process of automatically instrumenting an existing ROS system with monitors that intercept messages flowing between components (nodes). The monitor uses an oracle to decide if the intercepted messages obey/violate given formal properties. 
ROSMonitoring has several oracles, each built for a different formalism; to add support for a new formalism, a new oracle must be written manually. The oracle checks a message against the system's specification, and returns its verdict to the monitor. The system's specification is provided by the user.

\tikzset{every picture/.style={line width=0.75pt}} 

\begin{figure}[t]
\centering

\scalebox{0.6}{

\begin{tikzpicture}[x=0.75pt,y=0.75pt,yscale=-1,xscale=1]

\draw   (90,84) -- (172,84) -- (172,124) -- (90,124) -- cycle ;
\draw  [fill={rgb, 255:red, 224; green, 214; blue, 123 }  ,fill opacity=1 ] (46.1,68) -- (23,68) -- (23,22) -- (56,22) -- (56,58.1) -- cycle -- (46.1,68) ; \draw   (56,58.1) -- (48.08,60.08) -- (46.1,68) ;
\draw    (55,50) .. controls (82.3,48.05) and (69.67,66.06) .. (88.49,82.72) ;
\draw [shift={(90,84)}, rotate = 218.99] [fill={rgb, 255:red, 0; green, 0; blue, 0 }  ][line width=0.75]  [draw opacity=0] (8.93,-4.29) -- (0,0) -- (8.93,4.29) -- cycle    ;

\draw  [fill={rgb, 255:red, 108; green, 154; blue, 209 }  ,fill opacity=1 ] (317.1,127) -- (294,127) -- (294,81) -- (327,81) -- (327,117.1) -- cycle -- (317.1,127) ; \draw   (327,117.1) -- (319.08,119.08) -- (317.1,127) ;
\draw  [fill={rgb, 255:red, 108; green, 154; blue, 209 }  ,fill opacity=1 ] (337.1,147) -- (314,147) -- (314,101) -- (347,101) -- (347,137.1) -- cycle -- (337.1,147) ; \draw   (347,137.1) -- (339.08,139.08) -- (337.1,147) ;
\draw  [fill={rgb, 255:red, 108; green, 154; blue, 209 }  ,fill opacity=1 ] (354.1,175) -- (331,175) -- (331,129) -- (364,129) -- (364,165.1) -- cycle -- (354.1,175) ; \draw   (364,165.1) -- (356.08,167.08) -- (354.1,175) ;
\draw    (172,104) .. controls (199.36,118.93) and (224.25,141.77) .. (302.32,139.04) ;
\draw [shift={(303.5,139)}, rotate = 537.8299999999999] [fill={rgb, 255:red, 0; green, 0; blue, 0 }  ][line width=0.75]  [draw opacity=0] (8.93,-4.29) -- (0,0) -- (8.93,4.29) -- cycle    ;

\draw  [fill={rgb, 255:red, 108; green, 154; blue, 209 }  ,fill opacity=1 ] (304.1,236) -- (281,236) -- (281,190) -- (314,190) -- (314,226.1) -- cycle -- (304.1,236) ; \draw   (314,226.1) -- (306.08,228.08) -- (304.1,236) ;
\draw    (172,104) .. controls (197.37,116.94) and (178.69,193.23) .. (279.47,190.05) ;
\draw [shift={(281,190)}, rotate = 537.77] [fill={rgb, 255:red, 0; green, 0; blue, 0 }  ][line width=0.75]  [draw opacity=0] (8.93,-4.29) -- (0,0) -- (8.93,4.29) -- cycle    ;

\draw  [dash pattern={on 4.5pt off 4.5pt}] (230,34.5) -- (417.5,34.5) -- (417.5,263) -- (230,263) -- cycle ;
\draw    (315.5,201) .. controls (364.01,214.86) and (381.16,123.85) .. (444.56,121.06) ;
\draw [shift={(446.5,121)}, rotate = 539.12] [fill={rgb, 255:red, 0; green, 0; blue, 0 }  ][line width=0.75]  [draw opacity=0] (8.93,-4.29) -- (0,0) -- (8.93,4.29) -- cycle    ;

\draw  [fill={rgb, 255:red, 194; green, 108; blue, 214 }  ,fill opacity=1 ] (469.1,133) -- (446,133) -- (446,87) -- (479,87) -- (479,123.1) -- cycle -- (469.1,133) ; \draw   (479,123.1) -- (471.08,125.08) -- (469.1,133) ;
\draw   (467,192.5) .. controls (467,179.52) and (490.95,169) .. (520.5,169) .. controls (550.05,169) and (574,179.52) .. (574,192.5) .. controls (574,205.48) and (550.05,216) .. (520.5,216) .. controls (490.95,216) and (467,205.48) .. (467,192.5) -- cycle ;
\draw    (316.53,215.81) .. controls (382.37,241.21) and (409.05,167.76) .. (465.29,191.74) ;
\draw [shift={(467,192.5)}, rotate = 204.56] [fill={rgb, 255:red, 0; green, 0; blue, 0 }  ][line width=0.75]  [draw opacity=0] (8.93,-4.29) -- (0,0) -- (8.93,4.29) -- cycle    ;
\draw [shift={(314.5,215)}, rotate = 22.38] [fill={rgb, 255:red, 0; green, 0; blue, 0 }  ][line width=0.75]  [draw opacity=0] (8.93,-4.29) -- (0,0) -- (8.93,4.29) -- cycle    ;
\draw  [fill={rgb, 255:red, 119; green, 221; blue, 197 }  ,fill opacity=1 ] (458.1,274) -- (435,274) -- (435,228) -- (468,228) -- (468,264.1) -- cycle -- (458.1,274) ; \draw   (468,264.1) -- (460.08,266.08) -- (458.1,274) ;
\draw    (520.41,218.31) .. controls (519.83,242.69) and (531.75,257.34) .. (468.93,246.34) ;
\draw [shift={(467,246)}, rotate = 370.15] [fill={rgb, 255:red, 0; green, 0; blue, 0 }  ][line width=0.75]  [draw opacity=0] (8.93,-4.29) -- (0,0) -- (8.93,4.29) -- cycle    ;
\draw [shift={(520.5,216)}, rotate = 93.3] [fill={rgb, 255:red, 0; green, 0; blue, 0 }  ][line width=0.75]  [draw opacity=0] (8.93,-4.29) -- (0,0) -- (8.93,4.29) -- cycle    ;
\draw    (479,109) .. controls (527.02,122.72) and (471.32,147) .. (487.89,170.56) ;
\draw [shift={(489,172)}, rotate = 230.19] [fill={rgb, 255:red, 0; green, 0; blue, 0 }  ][line width=0.75]  [draw opacity=0] (8.93,-4.29) -- (0,0) -- (8.93,4.29) -- cycle    ;

\draw    (298.85,186.06) .. controls (285.58,145.79) and (303.42,154.29) .. (319.73,153.16) ;
\draw [shift={(321.5,153)}, rotate = 533.29] [fill={rgb, 255:red, 0; green, 0; blue, 0 }  ][line width=0.75]  [draw opacity=0] (8.93,-4.29) -- (0,0) -- (8.93,4.29) -- cycle    ;
\draw [shift={(299.5,188)}, rotate = 251.18] [fill={rgb, 255:red, 0; green, 0; blue, 0 }  ][line width=0.75]  [draw opacity=0] (8.93,-4.29) -- (0,0) -- (8.93,4.29) -- cycle    ;

\draw (131,104) node  [scale=1.0] [align=center] {instrument};
\draw (20,9) node [scale=1.0] [align=left] {config.yaml};
\draw (318,62) node [scale=1.0] [align=center] {nodes};
\draw (296,249) node [scale=1.0] [align=left] {monitor.py};
\draw (398,22) node  [scale=1.0] [align=left] {ROS};
\draw (462,74) node [scale=1.0] [align=left] {log.txt};
\draw (520.5,192.5) node  [scale=1.0] [align=center] {oracle};
\draw (450,215) node [scale=1.0] [align=left] {spec};
\draw (442,177) node [scale=1.0] [align=left] {online};
\draw (515,151) node [scale=1.0] [align=left] {offline};

\end{tikzpicture}

}
\caption{High-level overview of ROSMonitoring~\cite{Ferrando20a}. The configuration file guides the automatic instrumentation of the ROS system, where the nodes are modified (if necessary) and the monitor is synthesised. 
}
\label{fig:rosmon-pipeline}
\end{figure}
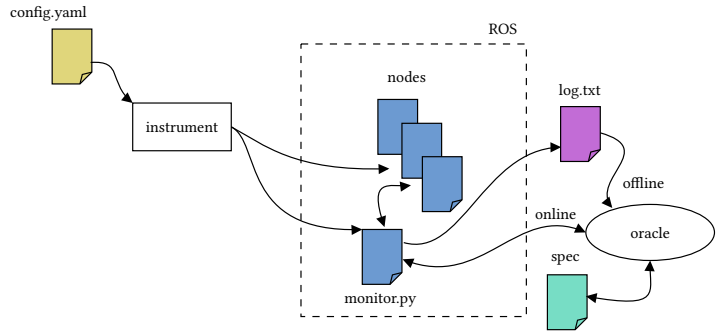

\section{Case Studies}
\label{sec:casestudies}
We outline four case studies that we were involved in where we elicited and formalised requirements using FRET. The  overarching aim of each of these case studies was different. Specifically, in the aircraft engine controller (\S\ref{subsec:enginecontroller}) we focused our efforts on eliciting and formalising requirements in conversation with an industrial partner (Collins Aerospace) \cite{farrellfretting2022}. In the mechanical lung ventilator (\S\ref{subsec:ventilator}) we formalised requirements that were supplied in the ABZ case study documentation with a view to developing a formal model of the system in Event-B \cite{abz2024}. For the inspection rover (\S\ref{subsec:inspectionrover}), we explored how we could integrate various verification artefacts into an assurance case and used a hazard analysis to define the requirements that were formalised in FRET and subsequently verified against models of the system using CoCoSim and Event-B \cite{bourbouhintegrating}. In the autonomous grasping case study (\S\ref{subsec:grasping}), we were tasked with verifying a system that already existed \cite{farrellformal2022}. Here, the developers had not formalised detailed requirements for verification so we used FRET to elicit and formalise the requirements, based on the components in an AADL model of the system. We modelled the code of the system using Dafny, and generated a suite of runtime monitors using ROSMonitoring that were able to identify requirement violations at run time.

\subsection{Case Study 1 - Aircraft Engine Controller}
\label{subsec:enginecontroller}

Our first case study is a software controller for a high-bypass civilian aircraft turbofan engine, based on existing controller designs~\cite{postlethwaitedigital1995,samardesign2010}. 
The example system was studied during the VALU3S~\cite{barbosavalu3s2020} project with direct input from our industrial partner (Collins Aerospace, Ireland).

The software controller is an example of a Full Authority Digital Engine Control (FADEC) system, which monitors and controls everything about the engine, including thrust control, fuel control, power management, health monitoring of the engine, thrust reverser control, etc.; using input from a variety of sensors.  The controller's high-level objectives are to manage engine thrust, regulate compressor pressure and speeds, and limit engine parameters to safe values. 
The controller must continue operating, keeping settling time, overshoot, and steady state errors within acceptable limits, while respecting the engine's operating limits in the presence of sensor faults, perturbation of system parameters and other low-probability hazards. The controller must also detect engine surge or stall and change mode to prevent these hazardous situations.

Our industrial partner defined 14 natural-language requirements. We used FRET during our requirements elicitation discussions and, based on the original 14 natural-language requirements and 20 associated test cases, we formalised 42 requirements using FRET \cite{farrellfretting2022}. We provide more detail in \S\ref{subsec:analysis} about how we formalised these requirements.

\subsection{Case Study 2 - Mechanical Lung Ventilator}
\label{subsec:ventilator}

\begin{figure}[t]
\centering
\includegraphics[width=\textwidth]{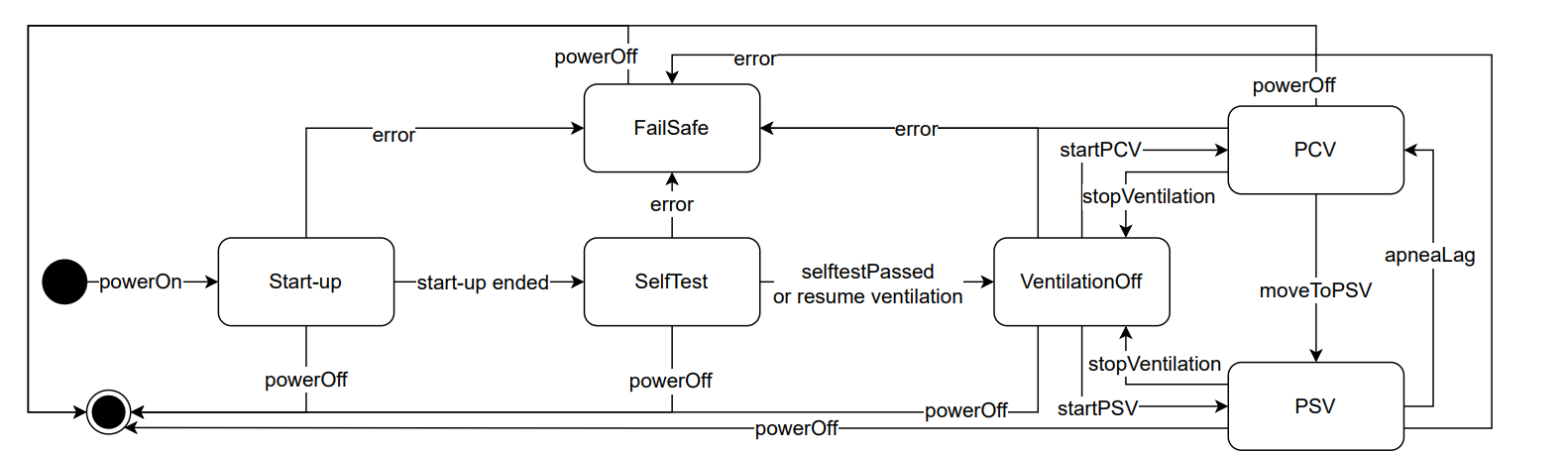}
\caption{Mechanical Lung Ventilator: The controller state machine is labelled as Fig 4.1 in the case study document.}
\label{fig:ventillatorarchitecture}
\end{figure}

Our next case study is a mechanical lung ventilator, from the case study track at the ABZ 2024 conference \cite{casestudyMLVSpec}.
During the COVID-19 pandemic, mechanical lung ventilators were essential medical equipment, supporting patients who were unable to breathe on their own \cite{bonivento2021mechanical,abba2021novel}. The system provides two ventilation modes: Pressure Controlled Ventilation (PCV) and Pressure Support Ventilation (PSV). The case study document includes a diagram of the system, shown in Fig.~\ref{fig:ventillatorarchitecture}. 

The case study document provides a large number of system requirements, split into: 58 Functional Requirements, 25 Values and Ranges, 58 Sensors and Interfaces, 107 GUI Requirements, 45 Controller Requirements, and 26 Alarms. 
The requirements contain two kinds of explicit dependency: requirements supported by `child' requirements, which are more like sub-clauses; and requirements that reference other requirements, as if using them as an external module\footnote{For examples of these dependencies see FUN6, FUN6\_1--6, and CONT4 in our requirements repository: \url{www.github.com/valu3s-mu/FRETISH-requirements}}. The documentation is unclear about the definition of these relationships.

For this case study, we used FRET to formalise 121 requirements, focusing on the functional (FUN) and controller (CONT) requirements. Following a requirements-driven approach, we then constructed an Event-B model using both the natural-language and FRET requirements as input to our process \cite{abz2024}.

\subsection{Case Study 3 - Inspection Rover}
\label{subsec:inspectionrover}

\begin{figure}[t]
    \centering
    \includegraphics[scale=0.58]{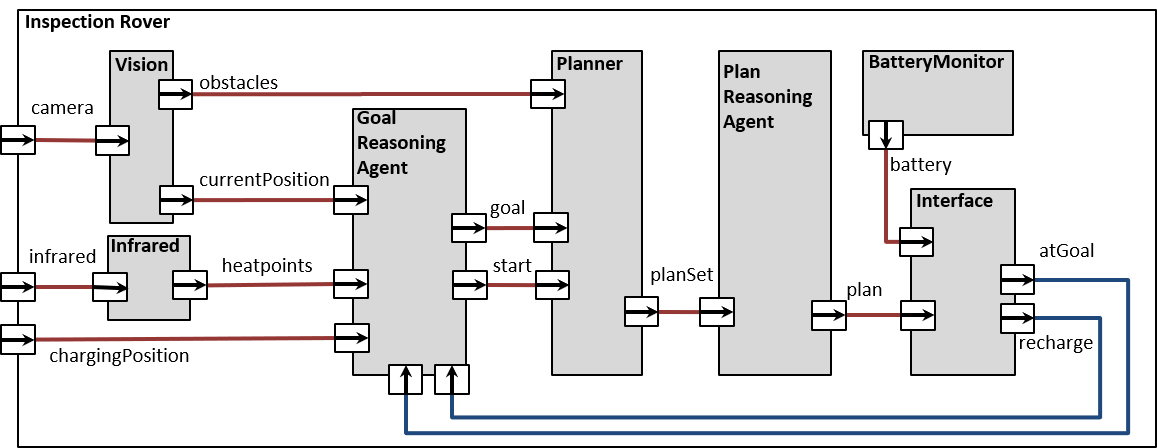}
    \caption{Inspection rover system architecture. The arrows show information flow.}
    \label{fig:inspectionrover}
\end{figure}

One of our authors elicited and formalised requirements for an inspection rover use case. The rover was tasked with navigating to points of interest on a 2D grid map of known size. Robotic systems tend to be quite modular, as embodied by the architecture in Fig. \ref{fig:inspectionrover}. Here, the \emph{Vision} system detects obstacles that the rover should avoid. 
The \emph{Infrared} component identifies grid locations that are hotter than expected (points of interest to be inspected). The autonomous \emph{Goal Reasoning Agent} chooses the hottest location as the next goal, unless the \emph{Battery Monitor} (via the \emph{Interface}) indicates that it must recharge. The \emph{Planner} computes obstacle-free plans for navigating to the goal. The autonomous \emph{Plan Reasoning Agent} selects the shortest plan. The \emph{Interface} translates the navigation actions of the plan into instructions for the hardware components and alerts the \emph{Goal Reasoning Agent} when it reaches the goal or must recharge. It is notified of the battery level by the dedicated \emph{Battery Monitor} component. 

We integrated formal verification and assurance approaches. First we used AdvoCATE~\cite{denney2012advocate} for assurance case development to conduct a hazard analysis on the model in Fig. \ref{fig:inspectionrover} and defined the system requirements. We formalised these 28 requirements using FRET and modelled the system in Event-B, Simulink and Lustre, subsequently verifying the requirements using Kind2 and Rodin.

\subsection{Case Study 4 - Autonomous Grasping}
\label{subsec:grasping}
\begin{figure}[t]
    \centering
    \includegraphics[scale = 0.056]{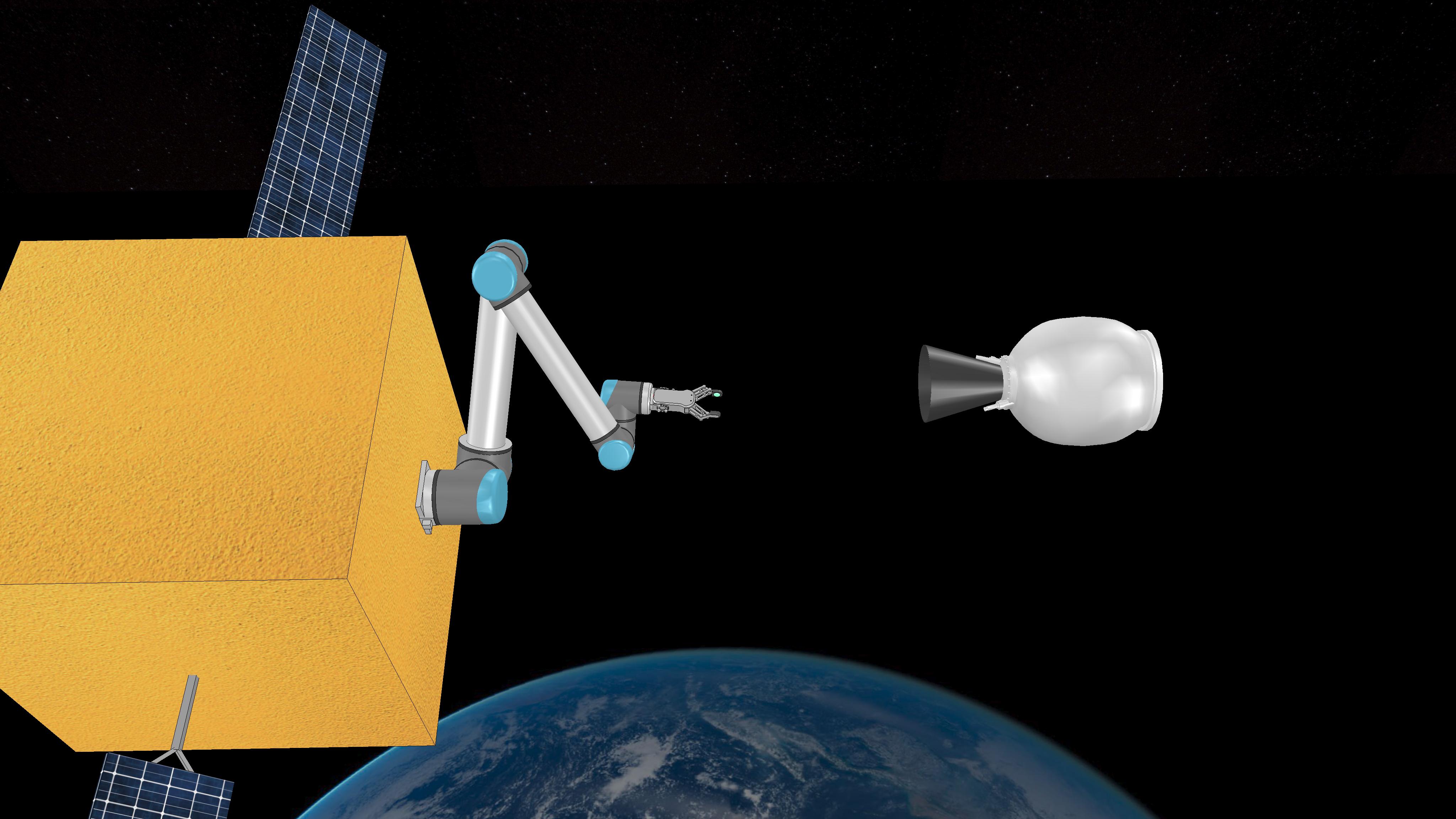}
    \hfill
    \includegraphics[scale = 0.032]{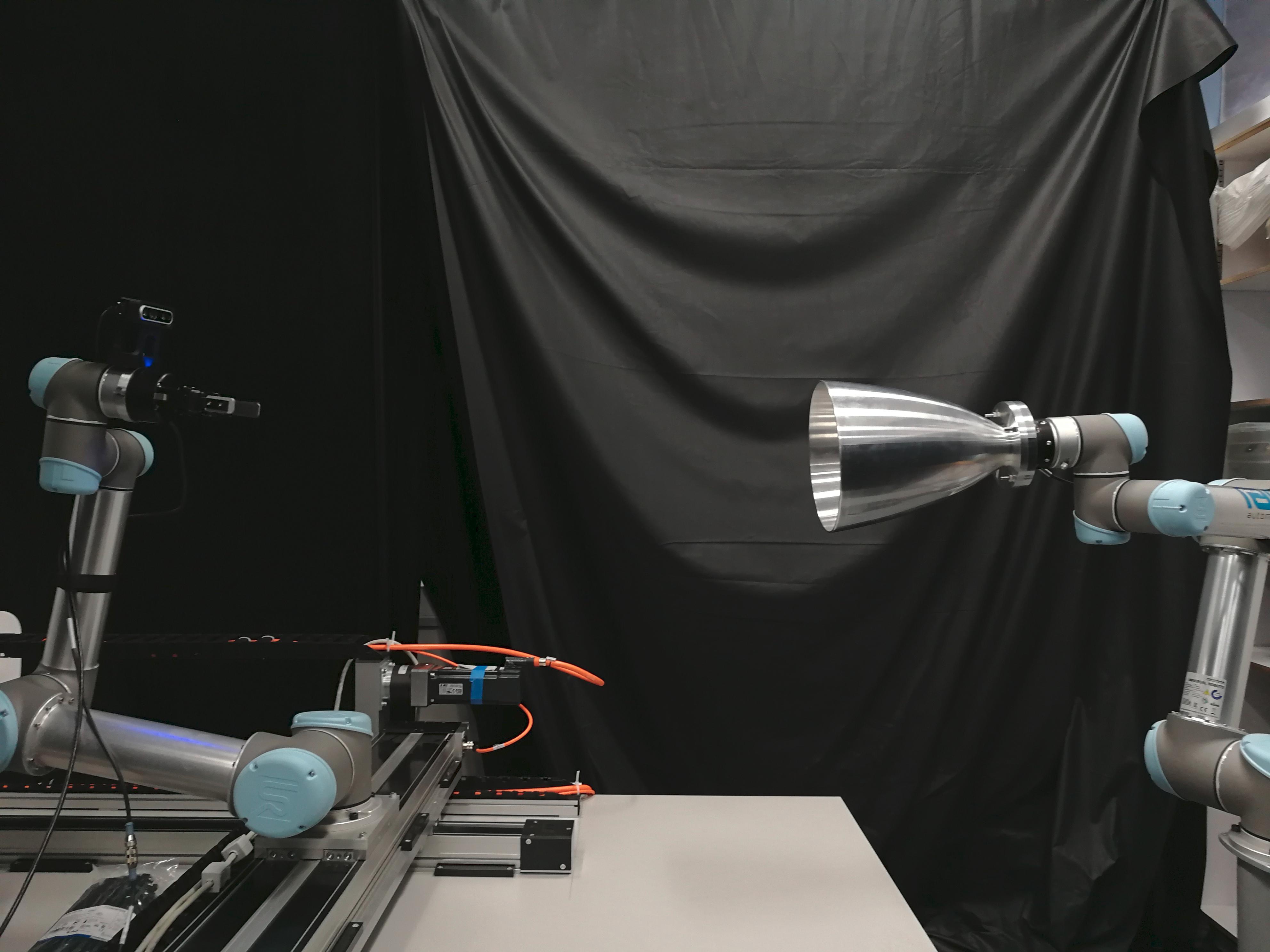}
    \caption{Simulation setup (left) and physical testbed for experimentation (right). The autonomous robotic arm (left of each image) must grasp the debris, an apogee kick motor (\texttt{AKM}) (right of each image) \cite{farrellformal2022}.}
    \label{fig:grasping}
\end{figure}

Another robotics use case focused on verifying an algorithm for autonomous grasping of spent rocket stages (apogee kick motors) for active debris removal in space \cite{mavrakis2019visually,farrellformal2022}. Fig. \ref{fig:grasping} illustrates the simulation environment and physical hardware testbed for this system. This differs to the previous case studies in that the task was to verify the system when the development process was almost complete, rather than using formal methods from the beginning. In order to elicit the most appropriate requirements for the system, an AADL \cite{feiler2006architecture} model of the system was developed. This prompted some reorganisation of the system but was helpful when assigning detailed requirements to specific system components.

Again, we used FRET for requirements elicitation. Different to the previous case studies, we used the Dafny program verifier \cite{leino2010dafny} for algorithm verification. To bridge the gap between system models and reality, we also generated runtime monitors for the system using ROSMonitoring \cite{ferrando2020rosmonitoring}. These monitors were deployed in both simulation and on the physical test bed.

With FRET, we formalised 20 requirements. The goal was to use formal methods (Dafny and ROSMonitoring) to verify our system. The \fretish representation was desirable as it more closely resembled formal properties than natural-language. The Dafny verification and runtime monitor generation was based on the \fretish requirements and FRET-generated LTL semantics. For example, one of the natural-language requirements stated: \textit{The position of the SV shall not be equal to the position of the TGT}. This was represented in \fretish as shown in Table \ref{tab:fretrequirements} (first requirement for the autonomous grasping use case). The corresponding runtime monitor is illustrated on the left of Fig. \ref{fig:monitorfig}.

\section{Formalising and Specifying Requirements}
\label{subsec:analysis}
In this section, we provide some metrics to characterise the case study requirements as expressed in \fretish. We discuss how we manually translated from \fretish to other formal specifications as mentioned in the previous sections. 

\subsection{Requirement Metrics}
\label{subsec:reqmetrics}

From \S\ref{sec:fretbg}, \fretish requirements are expressed using five fields: \fretishComponentsSmall{}.  
Table \ref{tab:fretrequirements} contains selected \fretish requirements from the case studies, showing how the requirements were expressed and how the various \fretish field options were instantiated. FRET generates a temporal logic semantics for requirements using an internal compositional approach. To achieve this, requirements are internally parsed into \emph{template keys}. Each template key is a tuple: [\emph{scope-option}, \emph{condition-option}, \emph{timing-option}] where \emph{scope-option} refers to the kind of scope (null, in, etc.) used in the \fretish requirement, \emph{condition-option} refers to the kind of condition (null, trigger, continual)\footnote{In previous versions of FRET, the trigger condition type was referred to as a \emph{regular} condition but the naming convention has been updated since then.}, and \emph{timing-option} refers to the kind of timing used (eventually, always, after n ticks, etc.).\footnote{If timing is omitted then  eventually timing is used in the generated LTL formulas.} Table \ref{tab:reqmetrics} contains the spread of the options used in the presented case studies. 
There are 10 timing options and 8 scope options available in FRET and we only discuss those that we used here but more detail is available in \cite{giannakopoulou2020formal}.

In \fretish, \Scope is generally used to denote modes of operation of the system. From Table \ref{tab:reqmetrics}, we observe that the \Scope field is rarely used in our case studies, outside of the Mechanical Lung Ventilator requirements. This is because these requirements described the interchange between various system modes (in Fig.~\ref{fig:ventillatorarchitecture}). By contrast, the Aircraft Engine Controller requirements only featured the `nominal' and `surge/stall prevention' operating modes, with no further detail of when the system should switch between them. This resulted in 38 of the 42 requirements having no scope specified. The natural-language requirements for the Inspection Rover and Autonomous Grasping system focused on distinct  components rather than system modes, so \Scope was not needed.

The case studies also differ in the use of the \ConditionF{} field. All requirements for the Aircraft Engine Controller had a \ConditionF{}, as each natural-language requirement specified a specific situation where it would apply (e.g. \textit{`Under sensor faults, while tracking pilot commands
'}). The Mechanical Lung Ventilator had a mix of requirements with/without condition options, due to interactions between many system components. The Autonomous Grasping and Inspection Rover case studies featured few of these trigger conditions, as these requirements typically specified rules that should apply for all operations of the concerned component.

We caveat our requirements analysis by saying that each of these requirements sets were developed using different FRET releases. The first case study was the inspection rover, followed by the autonomous grasping algorithm, then the aircraft engine controller, and finally the ABZ case study. During this evolution the syntax of FRET changed slightly to improve the user experience. For example the \Timing options in FRET were expanded as seen in the Mechanical Lung Ventilator, and a new kind of continual \conditionF{condition} was recently added.

From Table \ref{tab:reqmetrics}, we used many different kinds of \Timing{}. That said, we did not use the \timing{within}, \timing{before} or \timing{never} options (though all may not have been available in the releases used for the case studies). The \timing{never} timing can be represented using \timing{always} and logical negation. This approach was used in some of the case studies and an example is given in the first Autonomous Grasping requirement shown in Table \ref{tab:fretrequirements}. The most common timing options that we used were \timing{eventually} (or null), \timing{always}, and \timing{at the next timepoint}.

FRET does not support first-order temporal logic. To support the automatic translation from \fretish to CoCoSpec in the Inspection Rover case study, we used auxiliary variables in 7 requirements to represent quantifiers, and then instantiated these auxiliary variables since quantifiers are supported in CoCoSpec.

In FRET, users can specify parent-child relationships between requirements, allowing us to group related requirements together. In the aircraft engine controller we recorded how the original 14 natural language requirements were expanded to 42 through the addition of 28 child requirements. These child requirements came from a combination of discussion with  domain experts and examining the system test cases. In the Mechanical Lung Ventilator and Autonomous Grasping case studies, this relationship was used to promote traceability from the original document to the \fretish requirements where child (and grandchild) requirements represent sub-requirements in the original documentation.

\begin{table}[t]
    \centering
    
    \begin{tabularx}{\textwidth}{|c|X|}
    \hline
        \textbf{Case Study} &\textbf{ \fretish Requirement}  \\\hline \hline
        \textsf{Aircraft Engine Controller} & \scope{nominal} \conditionF{when (diff\_setNL\_observedNL > NLmax) if (pilotInput => surgeStallAvoidance)} \component{Controller} shall \timing{until (diff\_setNL\_observedNL < NLmin)} \responseF{(newMode = surgeStallPrevention)} \\ \hline 
         \textsf{\textsf{Aircraft Engine Controller}} & \conditionF{if ((sensorfaults) \& (trackingPilotCommands))} \component{Controller} shall \responseF{(controlObjectives)} \\ \hline \hline 
        \textsf{Mechanical Lung Ventilator} &  \scope{StartUpMode} \conditionF{when initDone} \component{Controller} shall \timing{at the next timepoint} \responseF{SelfTestMode} \\ \hline
         \textsf{Mechanical Lung  Ventilator} & \conditionF{when off} \component{System} shall \timing{after 15 minutes} \responseF{!resumeVentilation}  \\ \hline \hline 
        \textsf{Inspection Rover} & \component{Map\_Validator} shall \timing{immediately} \responseF{start = s0}\\ \hline
         \textsf{Inspection Rover} & \component{Rover} shall \timing{always} \responseF{battery > 0}  \\ \hline \hline
         \textsf{Autonomous Grasping} & \component{SV} shall \timing{always} \responseF{!(position(SV) = position(TGT))} \\ \hline
         \textsf{Autonomous Grasping} & \component{SV} shall  \responseF{(grasp(TGT, BGP) \& closer(SV, TGT))} \\ \hline \hline
    \end{tabularx}
    \caption{Example case study requirements showing what we can express in \fretish.}
    \label{tab:fretrequirements}
\end{table}

\begin{table}[t]

\textsf{Case Study 1: Aircraft Engine Controller}

\begin{tabularx}{\textwidth}{|c|X|}
\hline
\emph{scope-option} & null = 38, in = 4 \\\hline
\emph{condition-option} & trigger (regular) = 42 \\\hline
\emph{timing-option} &  null/eventually = 14, until = 28  \\\hline
parent-child &  28 child requirements were assigned a parent requirement   \\\hline
Total Requirements & 42, all expressed in \fretish.   \\\hline
\end{tabularx}\\

\textsf{Case Study 2: Mechanical Lung Ventilator}

\begin{tabularx}{\textwidth}{|c|X|}
\hline
\emph{scope-option} & null = 49, in = 70, before = 1, after =1 \\\hline
\emph{condition-option} & null = 51, trigger (regular) = 70 \\\hline
\emph{timing-option} & null/eventually =22, until =6, always=34, after=5, for =4,  next=50 \\\hline 
parent-child &  41 child requirements were assigned a parent requirement   \\\hline
Total Requirements & 142 but some could not be formalised in \fretish \\\hline
\end{tabularx}\\

\textsf{Case Study 3:Inspection Rover}

\begin{tabularx}{\textwidth}{|c|X|}
\hline
\emph{scope-option} & null = 28 \\\hline
\emph{condition-option} &  null = 27, trigger (regular) = 1 \\\hline
\emph{timing-option} & null/eventually = 13, always = 13, after =1, 
 immediately = 1\\\hline
 parent-child &  25 child requirements were assigned a parent requirement   \\\hline
Total Requirements & 28, all expressed in \fretish \\\hline
\end{tabularx}\\

\textsf{Case Study 4: Autonomous Grasping}

\begin{tabularx}{\textwidth}{|c|X|}
\hline
\emph{scope-option} & null = 20 \\\hline
\emph{condition-option} & null = 20 (5 used if then else in the response) \\\hline
\emph{timing-option} &  null/eventually = 17, always = 3  \\\hline
parent-child &  18 child requirements were assigned a parent requirement   \\\hline
Total Requirements & 20, all expressed in \fretish  \\\hline
\end{tabularx}\\

\caption{Metrics for \fretish requirements per case study, illustrating use of \Scope, \ConditionF, \Timing, number of  parent child  relationships and total requirements.}
\label{tab:reqmetrics}
\end{table}

\subsection{Specifying using FRETish and Event-B}
\label{subsec:fret2eb}

{
\begin{table}[t]
\textsf{Mechanical Lung Ventilator}
\begin{center}

\begin{tabular}{l|c|c|c|c}
        \hline
         \fretishbf{} \textbf{Req ID}& \textbf{Context(s)} & \textbf{Event(s)} & \textbf{Invariant(s)} & \textbf{Event-B File(s)}  \\ \hline\hline
         FUN4 &\checkmark &\checkmark & &\texttt{mac00, ctx00} \\
         FUN5 & &\checkmark & & \texttt{mac01} \\
         FUN5\_3 & &\checkmark & \checkmark & \texttt{mac01}\\
         FUN6 &\checkmark &\checkmark & & \texttt{mac00, mac01, ctx01} \\
         FUN6\_1--FUN6\_6 &\checkmark &\checkmark & &\texttt{mac01, ctx01} \\
         FUN7 & &\checkmark & &\texttt{mac01} \\
         FUN10 & &\checkmark & &\texttt{mac00} \\
         FUN10\_1 &\checkmark &\checkmark & &\texttt{mac01, ctx01} \\
         FUN10\_3--FUN10\_6 & &\checkmark & &\texttt{mac01} \\
         FUN23 & &\checkmark & &\texttt{mac01} \\
         FUN27 & &\checkmark & &\texttt{mac01} \\
         CONT1 &\checkmark &\checkmark & &\texttt{mac00, ctx00}\\
         CONT1\_1 & &\checkmark & &\texttt{mac01} \\
         CONT1\_3 & & &\checkmark &\texttt{mac01} \\
         CONT1\_6 & & &\checkmark &\texttt{mac01} \\
         CONT3 & &\checkmark & &\texttt{mac00} \\
         CONT4 & &\checkmark & &\texttt{mac00} \\
         CONT12 & &\checkmark & &\texttt{mac00, mac01} \\
         CONT18 &\checkmark &\checkmark & &\texttt{mac01, ctx01} \\
         CONT19 & &\checkmark & &\texttt{mac01} \\
         CONT38 & & &\checkmark &\texttt{mac01} \\
         CONT46 & &\checkmark & &\texttt{mac01} \\
         \hline              
 \end{tabular}
 \end{center}
 
\textsf{Inspection Rover}
\begin{center}

\begin{tabular}{l|c|c|c|c}
        \hline
         \fretishbf{} \textbf{Requirement ID}& \textbf{Context(s)} & \textbf{Event(s)} & \textbf{Invariant(s)} & \textbf{Event-B File(s)}  \\ \hline\hline
         R2.1 &\checkmark & & &\texttt{ctx0} \\      
         R2.4.1 &\checkmark & & &\texttt{ctx0} \\ 
         R2.4.3&\checkmark & & &\texttt{ctx0} \\  
         R2.4.4&\checkmark & & &\texttt{ctx0} \\  
         R2.5& & &\checkmark &\texttt{mac0} \\  
         R3.4& & &\checkmark &\texttt{mac1} \\\hline
 \end{tabular}
 \end{center}
    \caption{Moving between \fretish and \eventb contexts, events and invariants. \fretish requirements listed are detailed in \cite{abz2024} and \cite{bourbouhintegrating}. }
    \label{tab:fret2eb}
\end{table}}

\begin{table}[t]
\textsf{Autonomous Grasping}
\begin{center}
   
    \begin{tabularx}{\textwidth}{|l|c|c|c|X|}
    \hline
        \textbf{\fretish Req ID} & \texttt{requires} & \texttt{ensures} & \texttt{invariant} &\textbf{Dafny Method Name}   \\\hline \hline
        R1.3.2 & \checkmark & \checkmark &  &\texttt{imagepreprocessing}, \texttt{removeDepth}, \texttt{findoptimalgrasp}\\\hline
        R1.4 & &\checkmark & & \texttt{imagepreprocessing}\\\hline
        R1.5 & &\checkmark & & \texttt{findoptimalgrasp}\\\hline
        R1.6 & &\checkmark & & \texttt{findoptimalgrasp}\\\hline
        R1.5.1 & &\checkmark &\checkmark & \texttt{findoptimalgrasp}, \texttt{selectOptimalGrasp}\\\hline
        R1.5.2 &\checkmark & & & \texttt{selectOptimalGrasp}\\ \hline
    \end{tabularx}
\end{center}
    \caption{\fretish to Dafny. Note that we only indicate \texttt{invariant} here when the invariants used did not also support a post-condition.}
    \label{tab:fret2dfy}

\end{table}

Table \ref{tab:fret2eb} summarises how we represented our \fretish requirements in Event-B models for the mechanical lung ventilator and inspection rover case studies. In \cite{abz2024}, we provide a detailed discussion of how we used the \fretish requirements to construct the formal Event-B specification for the mechanical lung ventilator.

We encoded the \fretish requirements into \eventb in different ways, depending on what they specified. Some requirements were easily represented in a context, others became part of the behavioural event specifications, and some became invariant specifications. Since \eventb does not have direct support for temporal logic, we focused on the \fretish requirements and associated diagrammatic semantics, rather than attempting to represent temporal logic directly in \eventb. We found that some parent-child requirements were specified via event refinement in the \eventb model; e.g., by adding a timing parameter, resulting in  a \textit{superposition} refinement that constrains the event further \cite{abrial2007refinement}. However, not all of these relationships could be easily represented by refinement. The Ventilator requirements FUN5 and FUN5\_3 were both captured in the same \eventb machine as they seemed to be at the same level of abstraction. FUN5 described how start-up is initiated via a button press when the system is connected to the breathing circuit, air supply and power source. FUN5\_3 required the patient not be connected during start-up and self-test. This caused us to add functionality related to \texttt{SelfTestMode} which was not mentioned in FUN5.

\eventb does not have a native way to address timing properties, in contrast to \fretish, %
so we assumed the existence of a clock.
Using a clock variable also appears in related work that integrates timing constraints for Event-B~\cite{10.1007/11955757_13}. Due to time constraints, we did not explore the formalisation of all of the Ventilator requirements in \eventb. That said, we identified  over 60 of the remaining FUN and CONT requirements that should be formalisable in future refinement steps.

As mentioned in \S\ref{sec:casestudies}, the focus of our case studies differed and the two instances where we produced Event-B models were differently motivated. For the ventilator, we used the natural-language and \fretish requirements as input to our specification effort. For the inspection rover, we used the Event-B model to verify properties that could not be verified via Kind2 model-checking. Hence this model is focused only on a single component of the system, the planner. We modelled a simple planner and plan reasoning agent in Event-B and verified it against the relevant requirements. These focused mostly on producing obstacle free plans. We did not use the requirements to guide specification in this case, rather we added the requirements to a simple model. As such, the properties were represented as axioms in contexts and invariants in machines, rather than behavioural event specifications which were more common for the ventilator.

\subsection{Specification: FRET to Dafny}
\label{sec:specfret2dfy}
In the autonomous grasping case study, we modelled the system (originally a Python implementation) in Dafny. We added specification constructs to capture properties corresponding to the \fretish requirements, which are summarised in Table \ref{tab:fret2dfy}. Most of these were expressed as Dafny post-conditions (\texttt{ensures}) though some also were represented as pre-conditions (\texttt{requires}). Pre-conditions were normally used if the method in question relied on that property being true in order to function correctly, while methods with post-conditions produced an output that we showed was in accordance with the requirements. Wherever post-conditions were used, they were normally supported by various loop invariants as discussed in \cite{farrellformal2022}. The single invariant listed in Table \ref{tab:fret2dfy} is one that appeared in a method without an accompanying post-condition. 

\subsection{Specification: FRET to Temporal Logic}
\label{sec:specfret2ltl}
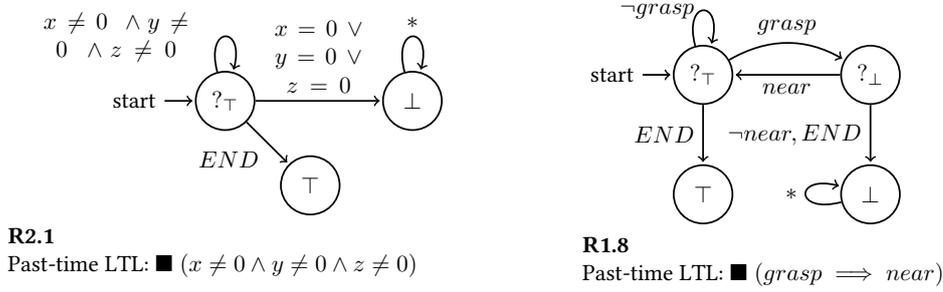
\begin{figure}[t]
\centering
\scalebox{0.88}{
\begin{minipage}{0.52\textwidth}
\begin{tikzpicture}[shorten >=1pt,node distance=1.8cm,on grid,auto] 
   \node[state,initial] (q_0)   {$?_\top$}; 
   \node[state] (q_1) [right=of q_0,xshift=1cm] {$\bot$};
   \node[state] (q_2) [below right=of q_0] {$\top$};
    \path[->] 
    (q_0) edge [align=center,text width=2cm] node {$x = 0 \;\lor$\\$y = 0 \;\lor$\\$z = 0$} (q_1)
          edge [align=center,loop above,text width=3cm] node [left]{$x \neq 0 \;\land y \neq 0 \;\land z \neq 0$} ()
          edge node [swap] {$END$} (q_2)
    (q_1) edge [loop above] node {$*$} ();
\end{tikzpicture}
\\ \textbf{R2.1}
\\ Past-time LTL: $\blacksquare\;(x \neq 0 \land y \neq 0 \land z\neq0)$
\end{minipage}
\qquad
\begin{minipage}{0.44\textwidth}
\begin{tikzpicture}[shorten >=1pt,node distance=1.8cm,on grid,auto] 
   \node[state,initial] (q_0)   {$?_\top$}; 
   \node[state] (q_1) [right=of q_0, xshift = 2em] {$?_\bot$};
   \node[state] (q_2) [below=of q_0] {$\top$};
   \node[state] (q_3) [right=of q_2, xshift = 2em] {$\bot$};
    \path[->] 
    (q_0) edge [bend left]  node {$grasp$} (q_1)
          edge [loop above] node [left]{$\lnot grasp$} ()
          edge node [swap] {$END$} (q_2)
    (q_1) edge node {$near$} (q_0)
          edge node [swap] {$\lnot near, END$} (q_3)
    (q_3) edge [loop left] node {$*$} ();
\end{tikzpicture}
\\ \textbf{R1.8}
\\ Past-time LTL: $\blacksquare\;(grasp \implies near)$
\end{minipage}}
    \caption{Runtime monitors for R2.2 (\emph{SV and TGT shall not be in the same position})  and R1.8 (\emph{the SVA shall capture the TGT at the BGP}) in the autonomous grasping case study. Note that $\blacksquare$ is the past-time version of the always operator.}
    \label{fig:monitorfig}
\end{figure}

In both the inspection rover and autonomous grasping case studies, we generated verification conditions for formal methods that supported temporal logics. For the inspection rover, we used FRET's analysis portal to generate CoCoSpec contracts that were verified against a Lustre model. This automatic step is supported by FRET \cite{mavridou2020bridging}, however some manual intervention was required to express properties with quantifiers for verification with Kind2. 
In the grasping case study, we used the \fretish requirements and Dafny specifications to manually construct runtime monitors. Fig. \ref{fig:monitorfig} contains examples of these monitors, represented as automata, and their past-time LTL equivalents. These were used to analyse the system both during simulation and on the physical test bed. The monitors were useful and identified bugs in the requirements that we were able to rectify.

\subsection{VerifyThis and DownSampling Point Clouds} 
\label{sec:verifythis}

Part of the autonomous grasping system was given as a challenge at VerifyThis 2022: specifically, the functionality of downsampling an input point cloud. Downsampling reduces the size of an input image before it is processed further. The resulting point cloud retains the overall geometric structure but has a reduced number of points. Techniques like this are common in signal/image processing and robotics. VerifyThis participants were provided with five   verification tasks: (1) memory safety, (2) termination, (3) the output point cloud is smaller or equal to the input point cloud (e.g. $size(pd) <= size(p)$), (4) The output point cloud is within the same range as the input point cloud (e.g. $boundingbox(pd) \ \ inside \ \ boundingbox(p)$), and (5) the output point cloud is a correct downsampled version of the input point cloud. The highest score received on this challenge was 2/5 and was achieved by three teams using Gobra, KeY and Frama-C. These teams primarily focused on verifying memory safety and termination for this challenge, some could specify but not verify the other properties. No team verified the correctness of tasks 4 and 5 for this challenge.

\section{Discussion}
\label{sec:discussion}

In this section, we outline and examine some specific discussion points. 

\paragraph{Communication:} Tools like FRET are invaluable when communicating with engineers and developers both in academia and industry. This was evidenced by the aircraft engine controller and autonomous grasping case studies. In both, the partners we were eliciting requirements from found that FRET's diagrammatic and natural-language semantics helped them to consider their system from different perspectives. This undoubtedly provided a bridge for these developers, providing traceability from natural-language to formalised requirements and verification conditions. This approach to demonstrating formal methods in a user-friendly way that leverages existing requirements engineering approaches encouraged our collaborators to consider using similar formal tools in the future.

\paragraph{Integrated Formal Methods:} Prior work  outlined the benefits that integrated formal methods can have in the robotics domain \cite{farrell2018robotics}. Both of our robotics case studies used multiple formal methods starting with requirements elicitation in FRET. In these case studies, some of the integration was via existing tool-supported translations (FRET to CoCoSpec) but the others were manual translations (FRET to Event-B, Dafny and ROSMonitoring). In the future, we hope that we can provide a more logically-founded and systematic approach to integrating these formal methods. Mathematical frameworks like UTP \cite{hoare1997unified} or institution theory \cite{goguen1992institutions} may have roles to play here. However, less formal approaches will also be beneficial including the definition of workflows to combine and use various methods alongside one another. 

\paragraph{Our Experience Writing \fretish:} In general, \fretish is intuitive and provides the user with a rich language to formalise requirements. However, we observed some limitations that may be considered in future extensions to the language and tool support. First, in both the aircraft engine controller and  inspection rover case studies, quantifiers in the language would have been incredibly useful. In the inspection rover we manually added quantifiers to the generated CoCoSpec contracts, and in the aircraft engine controller we used a specific naming convention to represent quantification over all sensors in the system. Quantifiers would have improved the overall expressiveness of \fretish in both cases and allowed us to write more elegant requirements. 
Another limitation of \fretish lies in the supported use of its \Scope field. 
A natural way to write a requirement that is active only in a single mode is `\scopeF{modeX} \component{System} shall \ldots', but this provides no in-built way to decribe a change to a new mode which is also represented as a \Scope{}. Of course, we can use a variable in the \ResponseF and/or \ConditionF{} to track the mode, for example \conditionF{if sys\_mode = modeX} \component{System} shall \timing{eventually} \responseF{sys\_mode = modeY}. 
The newer {\small\textcolor{Mahogany}{while}} keyword lets you write Boolean expressions to describe the mode, for example \scopeW{sys\_mode = modeX}, but this is less naturalistic to read than \scopeF{modeX}. 
It would be useful for \gls{fret} to have an in-built variable that tracks what mode the system is in, so that the user can use the more naturalistic {\small\textcolor{Mahogany}{in}} scope keyword and specify mode changes.

\paragraph{Our Experience Using FRET:}
Considering the translations that FRET supports, we found it difficult to produce and attach CoCoSpec contracts to our Simulink model of the Aircraft Engine Controller. This appeared to be due to the sheer size of our model, having come directly from our industry partner. We believe that improvements have been made to this translation based on our feedback, but ensuring that translations are scalable in general is something to always consider with any kind of formal method. There is benefit in integrating with tools like Simulink because they are widely-used in industry and this integration is a positive for encouraging uptake of FRET.

\paragraph{Requirement Management:} Throughout these case studies, we used FRET not only as an elicitation/formalisation tool but also as a way to manage the requirements that we were defining. We used the support that FRET provides for defining parent-child relationships between requirements, which allowed us to group related requirements together. Currently this parent-child link in FRET is informal in the sense that the user can assign a parent to a requirement without any formal/automatic checks that the requirements are related in some way. In future FRET developments it may be useful to formalise this relationship to allow the user to gradually refine requirements and provide a way to verify the traceability that we have documented between them. 

Some of our requirement sets were quite large (see Table \ref{tab:reqmetrics}). By assembling requirements in FRET, the various sets can be considered as software/system artefacts in their own right and thus managing them as development evolves is a non-trivial task. In other work, we have proposed refactoring techniques for \fretish requirements as a way to help with maintenance of large sets of requirements \cite{farrell2022towards}. This was motivated primarily by the aircraft engine controller requirement set, which was filled with repeated definitions that we recognised as potential issues for further development of the requirements and the system.

\section{Conclusion}
\label{sec:conclude}
This paper summarises our user-centric experience in expressing natural language requirements in \fretish, supporting the use of various existing formal methods. Our experience confirms FRET’s suitability as a framework for the elicitation, understanding, traceability of requirements. Our case studies cover those where the requirements were developed before the system was complete as well as those that retrospectively defined requirements to be used for verification after the system was complete. These two angles are discussed in \cite{farrellfretting2022,farrellformal2022}. FRET has changed over the course of these  case studies and improvements to the tool have directly resulted from our experiences. FRET continues to be improved by the team at NASA and we look forward to future developments that support users in specifying verifiable requirements.

\medskip
\noindent\textbf{\small{Acknowledgements.}} 
\label{sec:ack}
\small{We thank our co-authors and collaborators on each of the case studies: Stylianos Basagiannis, Georgios Giantamidis, Vassilios Tsachouridis, Hamza Bourbouh, Anastasia Mavridou, Irfan Sljivo, Guillaume Brat, Louise Dennis, Michael Fisher, Nikos Mavrakis, Angelo Ferrando, Yang Gao and Clare Dixon. }

\bibliography{main}
\bibliographystyle{plain}

\end{document}